# Sub-kHz-linewidth external-cavity laser (ECL) with $Si_3N_4$ resonator used as a tunable pump for a Kerr frequency comb

Pascal Maier,[1,2,*] Yung Chen,[1] Yilin Xu,[1,2] Yiyang Bao,[1,2] Matthias Blaicher,[1,2] Dimitri Geskus,[3] Ronald Dekker,[3] Junqiu Liu,[4] Philipp-Immanuel Dietrich,[1,2,5] Huanfa Peng,[1] Sebastian Randel,[1] Wolfgang Freude,[1] Tobias J. Kippenberg,[4,6] and Christian Koos[1,2,5,6,*]

[1] *Institute of Photonics and Quantum Electronics (IPQ), Karlsruhe Institute of Technology (KIT), Engesserstr. 5, 76131 Karlsruhe, Germany*
[2] *Institute of Microstructure Technology (IMT), Karlsruhe Institute of Technology (KIT), Hermann-von-Helmholtz-Platz 1, 76344 Eggenstein-Leopoldshafen, Germany*
[3] *LioniX International B.V., P.O. Box 456, 7500 AL Enschede, The Netherlands*
[4] *École Polytechnique Fédérale de Lausanne (EPFL), Rte Cantonale, 1015 Lausanne, Switzerland*
[5] *Vanguard Automation GmbH, Gablonzer Str. 10, 76185 Karlsruhe, Germany*
[6] *Deeplight S.A., Fondation EPFL Innovation Park – Bâtiment C, 1015 Lausanne, Switzerland*
* *pascal.maier@kit.edu; christian.koos@kit.edu*

**Abstract:** Combining optical gain in direct-bandgap III-V materials with tunable optical feedback offered by advanced photonic integrated circuits is key to chip-scale external-cavity lasers (ECL), offering wideband tunability along with low optical linewidths. External feedback circuits can be efficiently implemented using low-loss silicon nitride ($Si_3N_4$) waveguides, which do not suffer from two-photon absorption and can thus handle much higher power levels than conventional silicon photonics. However, co-integrating III-V-based gain elements with tunable external feedback circuits in chip-scale modules still represents a challenge, requiring either technologically demanding heterogeneous integration techniques or costly high-precision multi-chip assembly, often based on active alignment. In this work, we demonstrate $Si_3N_4$-based hybrid integrated ECL that exploit 3D-printed structures such as intra-cavity photonic wire bonds and facet-attached microlenses for low-loss optical coupling with relaxed alignment tolerances, thereby overcoming the need for active alignment while maintaining the full flexibility of multi-chip integration techniques. In a proof-of-concept experiment, we demonstrate an ECL offering a 90 nm tuning range (1480 nm–1570 nm) with on-chip output powers above 12 dBm and side-mode suppression ratios of up to 59 dB in the center of the tuning range. We achieve an intrinsic linewidth of 979 Hz, which is among the lowest values reported for comparable feedback architectures. The optical loss of the intra-cavity photonic wire bond between the III-V gain element and the $Si_3N_4$-based tunable feedback circuit amounts to approximately $(1.6 \pm 0.2)$ dB. We use the ECL as a tunable pump laser to generate a dissipative Kerr soliton frequency comb. To the best of our knowledge, our experiments represent the first demonstration of a single-soliton Kerr comb generated with a pump that is derived from a hybrid ECL.

## 1. Introduction

Tunable lasers are key building blocks of integrated optics. In this context, hybrid external-cavity lasers (ECL) are particularly interesting, opening the possibility to combine optical amplification in direct-bandgap III-V-based gain elements with tunable optical feedback offered by advanced passive photonic integrated circuits (PIC). Such feedback circuits can be efficiently implemented on the silicon photonic (SiP) platform [1]–[8]. Alternatively, silicon nitride ($Si_3N_4$) waveguides can be used for the feedback circuits [9]–[16], offering much lower linear and nonlinear losses than their SiP counterparts. This increases the intrinsic $Q$-factors of resonator-based feedback circuits and eliminates two-photon absorption (TPA) as a critical impairment at high power levels, thereby permitting greatly reduced phase noise and small linewidths [9], [17], [18]. However, the co-integration of III-V-based gain elements and tunable external-cavity circuits in chip-scale packages still represents a challenge. Specifically, heterogeneous integration based on transfer of InGaAsP epitaxial layers to pre-processed passive silicon or silicon nitride PIC paves a path towards high-density monolithic integration, but the underlying processes are still complex [19] and mostly limited to the SiP platform. This approach is hence mainly suited for high-volume applications that justify the associated technological overhead. Alternatively, III-V-dies and passive feedback circuits can first be processed on separate substrates and then be combined in a compact multi-chip module using a hybrid approach. This concept allows the components to be optimized and tested individually and facilitates thermal decoupling of the gain element from temperature-sensitive feedback circuits. However, assembly of hybrid photonic multi-chip modules crucially relies on high-precision alignment of the underlying optical dies, often requiring active alignment techniques, where the coupling efficiency is continuously measured and optimized during the assembly process [2], [8]–[12], [14]–[16]. This leads to limited fabrication throughput and often hinders cost-efficient scalability to high volumes.

In this paper, we demonstrate $Si_3N_4$-based hybrid integrated ECL that exploit photonic wire bonds (PWB) as intra-cavity coupling elements between reflective semiconductor optical amplifiers (RSOA) and the associated external-cavity circuits [20]. PWB are structured *in-situ* in a fully automated process with shapes adapted to the mode-field sizes and the positions of the chips at both ends, thus offering low coupling loss even for vastly dissimilar waveguide cross sections and limited placement accuracy of the underlying dies [21]. In a proof-of-concept experiment, we demonstrate an ECL offering a 90 nm tuning range (1480 nm–1570 nm) with on-chip output powers above 12 dBm and side-mode suppression ratios (SMSR) of up to 59 dB in the center of the tuning range. We achieve an intrinsic linewidth of 979 Hz, which is among the lowest values reported for comparable feedback architectures. The optical loss of the intra-cavity PWB between the III-V gain element and the $Si_3N_4$-based tunable feedback circuit amounts to approximately $(1.6 \pm 0.2)$ dB. At the output, the $Si_3N_4$-based tunable feedback circuit is coupled to an array of single-mode fibers using 3D-printed facet-attached microlenses (FaML) [22]–[25]. To demonstrate the performance and the versatility of the device, we use the ECL as narrow-linewidth tunable pump laser for generation of dissipative Kerr soliton (DKS) frequency combs in high-$Q$ $Si_3N_4$ microresonators. To the best of our knowledge, our experiments represent the first demonstration of a single-soliton Kerr comb generated with a pump that is derived from a hybrid integrated ECL.

## 2. Integrated hybrid external-cavity laser (ECL)

### 2.1 Device concept

The concept of a hybrid integrated ECL with 3D-printed intra-cavity photonic wire bond (PWB) and facet-attached microlenses (FaML) is shown in Fig. 1(a). The assembly consists of two dies on a common aluminum (Al) submount – an InP reflective semiconductor optical amplifier (RSOA) attached to a copper (Cu) heatsink and a $Si_3N_4$-based external-cavity circuit with a fiber array at the output. The device is controlled via an electronic circuit on a printed circuit board (PCB), which is attached to an Al base plate together with the submount. Figure 1(b) provides a top-view schematic of the InP RSOA and the $Si_3N_4$ chip, the latter comprising a Sagnac loop mirror with a Vernier pair of tunable racetrack resonators R1 and R2, a cavity phase tuner (CPT), and a tunable output coupler implemented as a Mach-Zehnder interferometer (MZI). The RSOA has a back facet with a high-reflectivity (HR) coating, while the front facet is angled at 9.0° and has an anti-reflection (AR) coating with respect to polymer. A PWB bridges the gap between the RSOA front facet and the edge coupler (EC) on the $Si_3N_4$ chip. This EC consists of a tapered waveguide (WG) which is oriented at an angle of 19.9° with respect to the facet normal. A side-view schematic of the module is provided in Fig. 1(c). Each end of the PWB is equipped with additional attachment structures that improve mechanical stability. Note that the PWB approach allows for connecting the two dies even though the emission angle of the RSOA output and the direction of the input EC of the $Si_3N_4$ chip are not matched, emphasizing the flexibility of the concept. The $Si_3N_4$ racetrack resonators, MZI, and CPT are tuned by thermal phase shifters. The CPT is used to adjust the cavity round-trip phase to an integer multiple of $2\pi$ at the wavelength corresponding to the maximum mirror reflectivity. The tunable MZI-based output coupler is used to set the ratio between the light extracted from the laser cavity and the feedback to the RSOA.

In the device implementation discussed in the following, the CPT is 1 mm long, and the Vernier pair of racetrack resonators have perimeters of 885.1 µm (R1) and 857.4 µm (R2). Four auxiliary waveguides WG 2…5 and the output WG 1 are routed to the $Si_3N_4$ chip edge and connected to tapered EC having a pitch of 127 µm. All five waveguide Ports 1…5 are coupled to a single-mode fiber (SMF) array (FA) via 3D-printed FaML.

### 2.2 Component characterization

To fully evaluate our integration concept, all components of the ECL are individually characterized prior to assembly. The assembly of fully characterized known-good components highlights one of the key advantages of our hybrid approach compared to other integration concepts. The following sections describe the measured performance of the RSOA and of the $Si_3N_4$ feedback circuit.

The C-band RSOA (Fraunhofer Heinrich-Hertz-Institut, Germany) is 700 µm long, has a back facet with a HR coating (90% reflectivity with respect to air) and an angled front facet (9°) with AR coating designed for emission into polymer ($n = 1.56$). The intensity distribution has an approximately circular shape with a 1/e²-diameter of 3.6 µm. The device is similar to the RSOA used in [1], and further details relating to the measurement techniques are described in the associated Supplementary Information. At a pump current of 100 mA, the measured maximum small-signal gain is 22 dB, and the saturated output power is $P_{sat} = 11.4$ dBm, both obtained at a wavelength of $\lambda = 1550$ nm. The RSOA characterization results are given in more detail in the Appendix B, see Fig. 5(a) and the associated description.

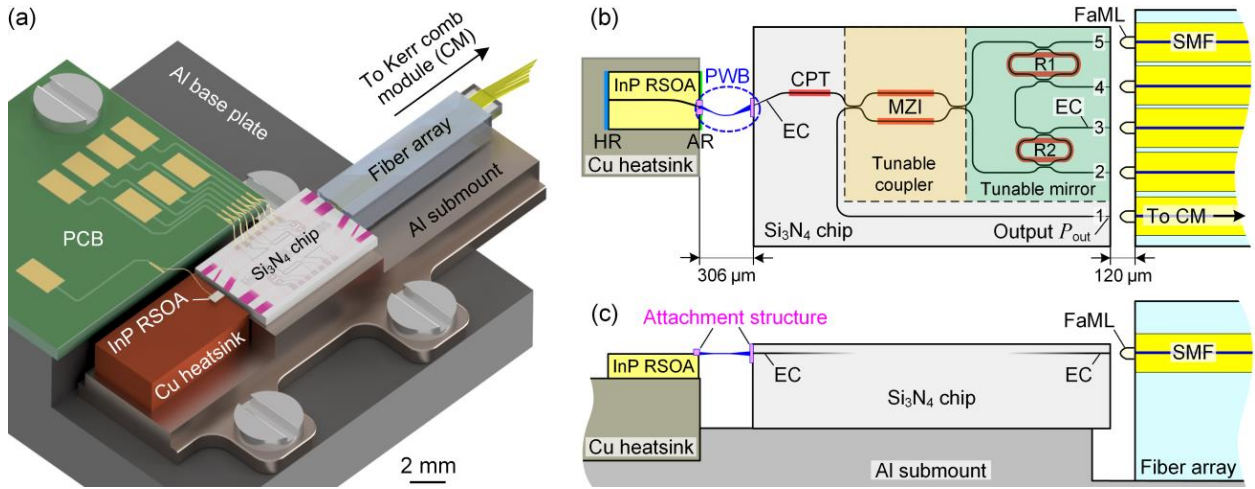

**Fig. 1.** Concept of a hybrid integrated ECL module with a 3D-printed intra-cavity photonic wire bond (PWB) and facet-attached microlenses (FaML). **(a)** The ECL assembly consists of two dies on a common aluminum (Al) submount – an InP reflective semiconductor optical amplifier (RSOA) attached to a copper (Cu) heatsink and a $Si_3N_4$-based external-cavity circuit with a fiber array at the output. The device is controlled via an electronic circuit on a printed circuit board (PCB), which is attached to an Al base plate together with the submount. The fiber-coupled ECL output can be connected to the Kerr comb module (CM), see Fig. 3(a). **(b)** Top-view schematic of the InP RSOA and the $Si_3N_4$ chip, the latter comprising a Sagnac loop mirror with a Vernier pair of tunable racetrack resonators R1 and R2, a cavity phase tuner (CPT), and a tunable output coupler implemented as a Mach-Zehnder interferometer (MZI). The RSOA has a back facet with a high-reflectivity (HR) coating, while the front facet is angled at 9.0° and has an anti-reflection (AR) coating with respect to polymer. The intra-cavity PWB bridges the gap of 306 µm between the RSOA front facet and the edge coupler (EC) on the $Si_3N_4$ chip. This EC consists of a tapered waveguide (WG), which is oriented at an angle of 19.9° with respect to the facet normal. Note that the PWB approach allows for connecting the two dies even though the emission direction of the on-chip waveguides at its ends are not matched, emphasizing the flexibility of the concept. In our device implementation, the CPT is 1 mm long and the Vernier pair of racetrack resonators have perimeters of 885.1 µm (R1) and 857.4 µm (R2). Four auxiliary waveguides WG 2…5 and the output WG 1 are routed to the $Si_3N_4$ chip edge and connected to associated tapered EC having a pitch of 127 µm. All five waveguide Ports 1…5 are coupled to a single-mode fiber (SMF) array via 3D-printed FaML. **(c)** Side-view schematic of the module. Each end of the PWB is equipped with additional attachment structures that improve mechanical stability. A step on the Al submount is used to align the height of the RSOA chip surface to that of the WG layer containing the EC on the $Si_3N_4$ side.

Frequency-selective optical feedback is provided by an ECL filter chip comprising a Vernier pair of racetrack resonators on the $Si_3N_4$ chip (TriPleX®, LioniX International B.V., The Netherlands), Fig. 1(b). The underlying waveguides are based on a symmetric double-stripe geometry consisting of a pair of stacked $Si_3N_4$ layers buried in a $SiO_2$ cladding – a standard structural element of the TriPleX® platform [26], [27]. The low propagation loss of less than 0.1 dB/cm is achieved with a width of the two $Si_3N_4$ stripes of 1.1 µm and a height of only 175 nm for the top and 75 nm for the bottom $Si_3N_4$ stripe, respectively. Both racetrack resonators have equal WG cross sections, leading to identical propagation constants and slightly different free spectral ranges (FSR) achieved by slight differences of the perimeters. The individual racetrack resonators are characterized by power transmission measurements in a loop-back configuration through the two auxiliary waveguide Ports 3 and 4, see Fig. 1(b). The measured transmission can be fit to a model to estimate the resonator parameters such as the loaded $Q$-factors and the FSR, see Appendix A and [1] for more details. At a wavelength of 1525 nm, corresponding to the center of our tuning range, we extract FSR of 195.7 GHz and 202.0 GHz for R1 and R2, respectively, along with approximately equal loaded $Q$-factors of 20 000. With the parameters obtained from the fit, we can calculate a 0.1 dB on-chip reflection loss of the tunable mirror, measured at the top of the reflection peak in case a resonance of R1 is perfectly aligned with a corresponding resonance of R2.

### 2.3 Module assembly

In a first step of module assembly, the RSOA is glued to a copper (Cu) heatsink with an electrically conductive adhesive (EPO-TEK® H20E, Epoxy Technology Inc., USA). The Cu heatsink with the RSOA and the $Si_3N_4$ chip are then coarsely aligned to each other and glued to an Al submount, see Fig. 1(a). A step on the Al submount is used to adjust the height of the RSOA chip surface to that of the WG layer containing the EC on the $Si_3N_4$ external-cavity chip. The PWB trajectory depends on the exact positions and emission directions of the RSOA and $Si_3N_4$ EC facets, see Fig. 1(b). The EC on the external-cavity chip rely on spot-size converters (SSC) that are implemented by simultaneously tapering the thickness of the top $Si_3N_4$ layer as well as the width of the two stacked $Si_3N_4$ stripes, while maintaining a constant thickness of the intermediate $SiO_2$ layer. In the EC used for our devices, the thickness of the top $Si_3N_4$ layer is tapered down from the usual 175 nm to zero close to the end-facets, effectively only leaving the bottom $Si_3N_4$ stripe with 75 nm thickness. At the input of the chip, the $Si_3N_4$ stripes start with an initial width of 2 µm at the facet, which is reduced to the standard width of 1.1 µm at the end of the taper section. This leads to a measured elliptical mode-field with 1/e²-width of 7.2 µm in the horizontal and 4.6 µm in the vertical direction, and the rectangular cross-section of the PWB is chosen to have a width of 8.0 µm and a height of 5.0 µm for best coupling to the mode-field size at the chip facet. In our implementation, the PWB bridges a gap of 306 µm between the two dies. The PWB on the RSOA side comprises a taper with an initial cross-section of 4.0 µm × 4.0 µm, matched to the mode-field size at the RSOA facet. The PWB cross-section is then reduced to the usual 2.4 µm × 2.0 µm. At the facet of the Si$_3$N$_4$ chip, the PWB is up-tapered to the above-mentioned final cross-section of 8.0 µm × 5.0 µm connecting to the on-chip Si$_3$N$_4$ EC. Between the tapers, the PWB maintains the constant cross-section of 2.4 µm × 2.0 µm, while the trajectory describes an arc to smoothly connect to the edge-coupled on-chip WG at either end, which feature notably different directions. Printed attachment structures on both the RSOA and the Si$_3$N$_4$ chip facets improve the mechanical stability of the PWB, see Fig. 1(c).

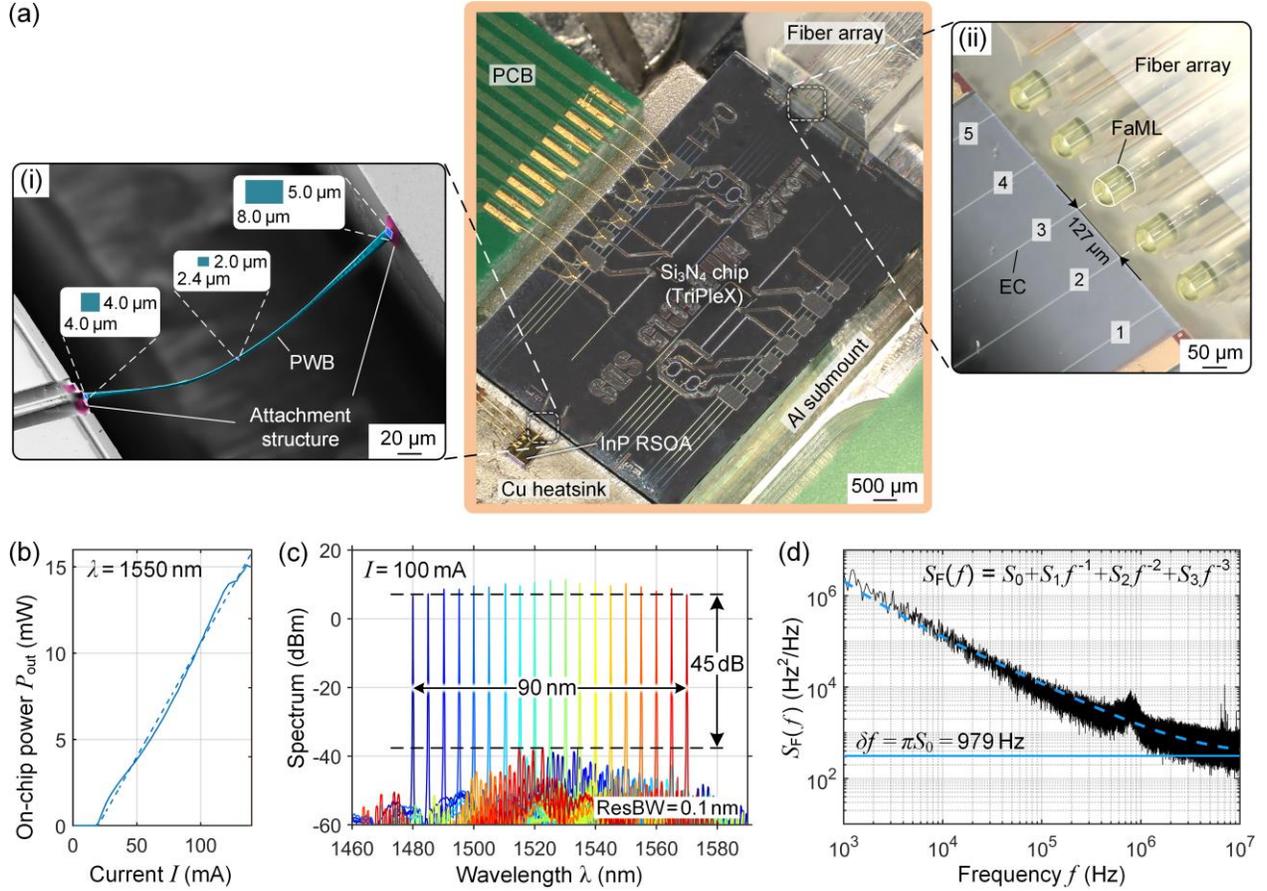

**Fig. 2.** Implementation of ECL module and performance characterization. **(a)** Microscope image of the assembled ECL module according to Fig. 1(a). Inset (i) shows a scanning-electron microscope (SEM) image of the false-colored PWB between the RSOA and Si$_3$N$_4$ (TriPleX®) chip. The various cross-sections at different positions along the PWB are marked. Inset (ii) displays a microscope image of the five FaML on the FA having a pitch of 127 µm and facing the corresponding EC of the waveguide Ports 1…5 on the Si$_3$N$_4$ chip. **(b)** On-chip output power $P_{out}$ vs. injection current $I$ for lasing at 1550 nm, leading to a threshold current of 19 mA and an average slope efficiency of 132 mW/A (fit by dashed straight line). **(c)** Superimposed ECL lasing spectra with tuning steps of 5 nm within the tuning range (1480…1570) nm for $I = 100$ mA. The maximum on-chip output power is 12 dBm near a wavelength of 1525 nm. The side-mode suppression ratio (SMSR) is at least 45 dB and can reach 59 dB in the center of the tuning range, see Fig. 5(b) below. **(d)** FM noise spectrum with fit (blue dashed line). The intrinsic (Lorentzian) linewidth is $\delta f = \pi S_0 = 979$ Hz (fit by solid blue line).

Fiber-chip coupling at the output of the Si$_3$N$_4$ chip relies on five identical FaML, printed to the facets of the FA. These FaML have been designed for the specific EC used at the output facets of the external-cavity chip towards the fiber array (waveguide Ports 1 … 5). The EC again rely on a pair of Si$_3$N$_4$ stripes, which are now down-tapered from the standard width of 1.1 µm to a final width of 0.8 µm at the facet, with the bottom Si$_3$N$_4$ layer keeping its constant thickness of 75 nm while the thickness of the upper layer is reduced from the initial value of 175 nm to zero at the facet. The resulting mode-field was measured to be 7.5 µm in both directions. The FaML are 70 µm long and offer a working distance of 50 µm between the apices of the lenses and the Si$_3$N$_4$ WG facets. The distance between the FA and the Si$_3$N$_4$ chip facet hence amounts to 120 µm, see Fig. 1(b).

Both the PWB and the FaML are made from a negative-tone photoresist (VanCoreA, Vanguard Automation GmbH, Germany; refractive index $n = 1.53$) by *in-situ* multi-photon lithography. The fabricated PWB and FaML are developed in propylene-glycol-methyl-etheracetate (PGMEA), flushed with isopropanol, and finally blow-dried. The FaML on the FA are 3D-printed in a separate step for subsequent assembly with a custom pick-and-place machine [25]. Figure 2(a) shows a microscope image of the fully assembled ECL module. Inset (i) displays a scanning-electron microscope (SEM) image of the false-colored PWB along with the dimensions of the various PWB cross-sections. Inset (ii) shows a microscope image of the five FaML on the FA facing the corresponding edge-coupled waveguide Ports 1…5 on the $Si_3N_4$ chip.

*2.4 Module characterization*

The ECL wavelength is selected by aligning the two racetrack resonators to a common resonance, which may be detected through the two auxiliary waveguide Ports 3 and 4 in a loop-back configuration. After the alignment of the resonances, the cavity phase and the tunable coupler are adjusted for maximum on-chip output power $P_{out}$, see Fig. 1(b). From the power-current characteristics (*P-I*-curve) in Fig. 2(b) measured at a wavelength $\lambda = 1550\,\text{nm}$, we extract a threshold current of 19 mA and an average slope efficiency of 132 mW/A. We further estimate a PWB loss of $(1.6 \pm 0.2)\,\text{dB}$ by measuring the ECL emission power $P_{out}$ and by comparing it to the measured saturated output power $P_{sat}$ of the RSOA at the same wavelength $\lambda = 1550\,\text{nm}$ and at the same injection current $I$. Details relating to the estimation of the PWB losses can be found in the Appendix B. The measured PWB loss is on par with previously reported coupling losses of $(1.6 \pm 1.0)\,\text{dB}$ between RSOA and $Si_3N_4$-based PIC using passive alignment [28], outperforming many previous demonstrations featuring typical RSOA-to-chip insertion losses of $(2…3)\,\text{dB}$ [11], [13], [15], [16], [29]. Butt-coupling losses below 1 dB are also feasible, but the associated devices require carefully laid-out SSC on the passive chips [2], [3] or inverse tapers on both the active and passive chips [7], [30], usually in conjunction with active alignment techniques for precise positioning.

Figure 2(c) depicts superimposed ECL lasing spectra with tuning steps of 5 nm within the tuning range $(1480…1570)\,\text{nm}$ for a fixed injection current of $I = 100\,\text{mA}$. A maximum on-chip output power $P_{out}$ of 12 dBm is measured at wavelengths near the center of the tuning range $(\lambda_c = 1525\,\text{nm})$, close to the center wavelength of the RSOA gain spectrum. The SMSR is at least 45 dB over the whole tuning range and reaches 59 dB in the center of the tuning range, see Fig. 5(b) below.

We further measure the phase noise of the emitted laser line using a frequency discriminator (LWA-1k 1550, HighFinesse GmbH, Germany). The measurements are carried out for an emission wavelength close to 1550 nm, where we re-adjust the tunable output coupler for minimum intrinsic linewidth. The associated FM noise spectrum is shown in Fig. 2(d). The intrinsic (Lorentzian) linewidth is obtained by first fitting a model function of the form $S_F(f) = S_0 + S_1 f^{-1} + S_2 f^{-2} + S_3 f^{-3}$ to the measured FM noise spectrum (blue dashed line). This leads to $S_0 = 311\,\text{Hz}^2/\text{Hz}$, corresponding to an intrinsic linewidth of $\delta f = \pi S_0 = 979\,\text{Hz}$ (solid blue line), which is among the lowest values reported for similar feedback architectures [2]–[4], [6], [9]–[12]. The low phase noise results from the large effective optical cavity length of $L_{eff} = 48.7\,\text{mm}$ at the center of the emission range $(\lambda_c = 1525\,\text{nm})$, the low passive WG losses (0.1 dB/cm), and the small RSOA-to-chip coupling loss (1.6 dB). The effective optical cavity length $L_{eff}$ accounts for the group delay of all cavity elements including the RSOA waveguide, the PWB, the on-chip $Si_3N_4$ waveguides as well as the highly dispersive racetrack resonators R1 and R2 and is hence much larger than the geometrical cavity length. Minimizing losses in the extended laser cavity, which are dictated mainly by the RSOA-to-chip coupling loss, is essential to obtain low laser linewidths [29]. Our results are on par with lately demonstrated linewidths of 895 Hz [12], achieved with actively aligned $Si_3N_4$-based external cavity circuits featuring only a pair of Vernier-type ring resonators. Note that previously reported demonstrations of such devices usually reached values of the order of 2 kHz [9], [10]. Lower intrinsic linewidths may be achieved by using external cavity circuits with three or more ring resonators: Using shallow-etched low-loss SiP rib WG in combination with three-ring external cavity circuits, heterogeneously integrated ECL with linewidths down to 95 Hz have been demonstrated [5]. The linewidths of three-ring ECL can be further reduced down to 40 Hz by using $Si_3N_4$-based external cavity circuits with large optical cavity lengths of the order of 0.5 m [14]. Note that feedback circuits based on standard SiP strip WG generally lead to larger intrinsic linewidths of, e.g., 12 kHz, demonstrated with a hybrid integrated ECL based on a pair of Vernier-type SiP ring resonators in combination with a Mach-Zehnder-type delay interferometer that additionally suppresses Vernier side-lobes in the filter transfer function [8].

## 3. Kerr comb generation

To demonstrate the versatility of our hybrid ECL, we use the device as a precisely tunable pump laser to generate dissipative Kerr soliton (DKS) frequency combs in $Si_3N_4$ microresonators. Figure 3(a) shows a schematic of the associated experimental setup: The ECL fiber output is connected to an optical isolator, followed by an erbium-doped fiber amplifier (EDFA) and a polarization controller (PC). The light is then fed via a PWB to a separately packaged comb module (CM), consisting of a ring resonator (R3, diameter 1286 µm, free spectral range 35.4 GHz, $Q \approx 15 \times 10^6$), which was fabricated using the photonic Damascene reflow process [31], [32]. More details on the characterization procedures and results for the $Si_3N_4$-based chip used for the CM are given in Appendix C. A microscope image of the CM is shown at the bottom of Fig. 3(a) along with a top-view schematic of the $Si_3N_4$ chip. Besides the ring resonator R3 of interest, the chip contains two additional rings that were not used in our experiments. The output of the ring resonator R3 is connected to a subsequent FA by a PWB. The input and output FA as well as the $Si_3N_4$ chip are fixed to a common Al submount.

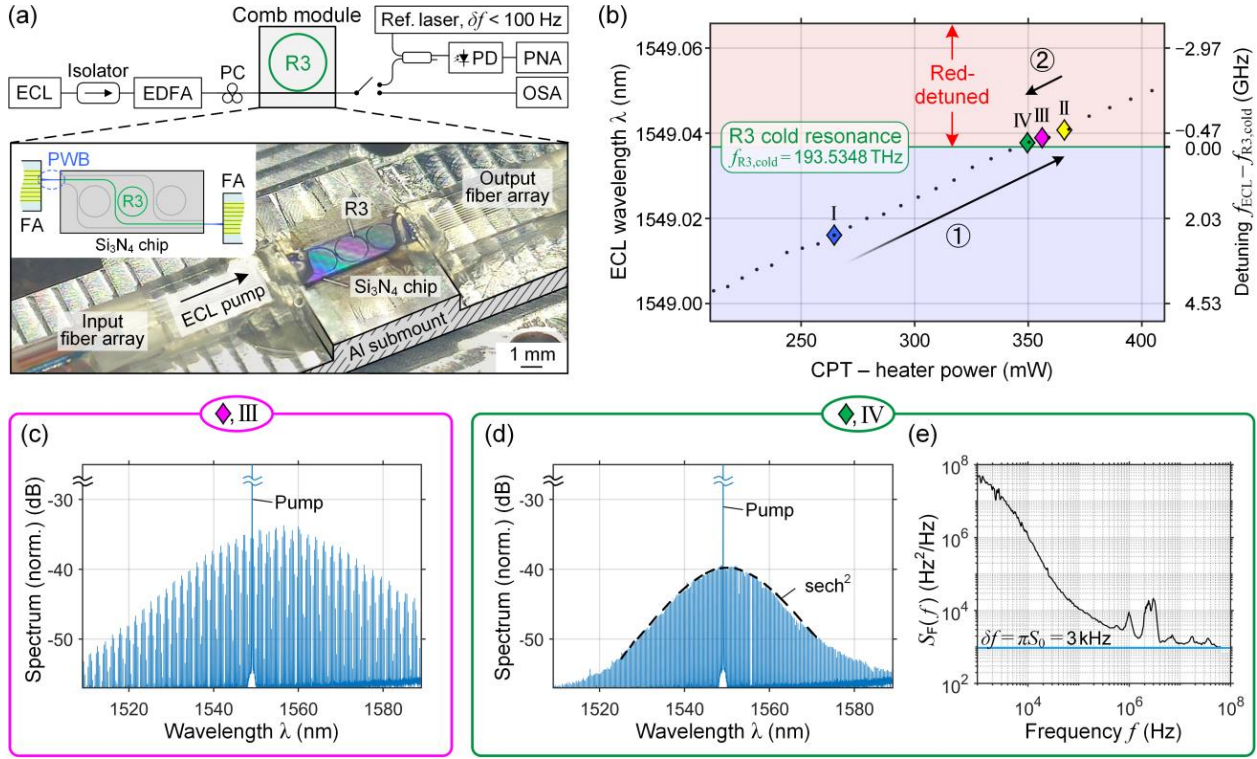

**Fig. 3.** Dissipative Kerr soliton (DKS) frequency comb generation in a high-$Q$ $Si_3N_4$ microresonator (R3) using the ECL module as a precisely tunable pump laser. **(a)** Schematic of the associated experimental setup: The ECL fiber output is connected to an optical isolator followed by an erbium-doped fiber amplifier (EDFA) and a polarization controller (PC). The light is then fed via a PWB to a separately packaged comb module (CM), consisting of a ring resonator (R3, diameter 1286 µm, free spectral range 35.4 GHz, $Q \approx 15 \times 10^6$), which was fabricated using the photonic Damascene process. A microscope image of the CM is shown in the bottom part of the figure along with a top-view schematic of the $Si_3N_4$ chip. Besides the ring resonator R3 of interest, the chip contains two additional rings that were not used in our experiments. The output of the ring resonator R3 is connected to a subsequent FA by a second PWB. The input and output FA as well as the $Si_3N_4$ chip are fixed to a common Al submount. The output of the CM is either monitored on an optical spectrum analyzer (OSA) or beat with a highly stable reference laser tone. The photocurrent at the intermediate frequency is sent to a phase-noise analyzer (PNA). **(b)** Start-up process for the comb source, relying on precise tuning of the ECL emission wavelength. The left axis indicates the absolute wavelength of the ECL, whereas the axis on the right shows the associated detuning $f_{ECL} - f_{R3,cold}$ of the ECL frequency $f_{ECL}$ with respect to the cold resonance of R3 $f_{R3,cold} = 193.5348\,THz$ (horizontal green line). Initially, the ECL wavelength (♦, I) is blue-detuned with respect to the cold resonance of R3. Then the ECL emission wavelength is increased by ramping up the CPT heater power, see black dots in Fig. 3(b) and Arrow ①. Upon reaching the red-detuned regime, $f_{ECL} < f_{R3,cold}$, a multi-soliton state can be observed (♦, II). Subsequently, the ECL pump is tuned backwards (Arrow ②), thereby reducing the number of solitons circulating in the cavity (♦, III), until finally, a single-soliton state is reached at an ECL wavelength slightly above the cold resonance of R3 (♦, IV). **(c)** Spectrum of an exemplary multi-soliton comb, characterized by spectral fringes. The multi-soliton state is indicated by a pink diamond (♦, III) in (b). **(d)** Spectrum of single-soliton comb, characterized by a smooth spectral envelope, which can be well approximated by the typical $sech^2$-characteristic. The single-soliton state is indicated by a green diamond (♦, IV) in (b). **(e)** FM noise spectrum of a typical single-soliton comb line. We estimate an intrinsic linewidth of $\delta f = 3\,kHz$.

The start-up process for the comb source is illustrated in Fig. 3(b) [33], [34]. The left axis indicates the absolute wavelength of the ECL, whereas the axis on the right shows the associated detuning $f_{ECL} - f_{R3,cold}$ of the ECL frequency $f_{ECL}$ with respect to the cold resonance of R3 $f_{R3,cold} = 193.5348\,THz$ (horizontal green line). Initially, the ECL wavelength (♦, I) is blue-detuned with respect to the cold resonance of R3. Then the ECL emission wavelength is increased by ramping up the CPT heater power, see black dots in Fig. 3(b) and Arrow ①. Upon reaching the red-detuned regime, $f_{ECL} < f_{R3,cold}$, a multi-soliton state can be observed (♦, II). Subsequently, the ECL pump is tuned backwards [33], [34] (Arrow ②), thereby reducing the number of solitons circulating in the cavity (♦, III), until finally, a single-soliton state is reached at an ECL wavelength slightly above the cold resonance of R3 (♦, IV). For the states (♦, III) and (♦, IV), we record the spectrum using an optical spectrum analyzer (OSA), Fig. 3(c) and Fig. 3(d). The spectral envelope of the single-soliton state (♦, IV) is smooth and can be well approximated by the typical $sech^2$-characteristic, Fig. 3(d), whereas the multi-soliton state (♦, III) is characterized by spectral fringes, Fig. 3(c). To the best of our knowledge, our experiments represent the first demonstration of a single-soliton Kerr comb generated with a pump that is derived from a hybrid integrated ECL.

We also measure the FM noise spectrum of a typical single-soliton comb line, Fig. 3(e), by beating it with a highly stable reference laser tone (Koheras X15, NKT Photonics A/S, Denmark; intrinsic linewidth $\delta f < 100\,\text{Hz}$) on a photodetector (PD). The photocurrent at the intermediate frequency is sent to a phase-noise analyzer (PNA, FSWP50, Rohde & Schwarz GmbH & Co. KG, Germany). From the recorded FM noise spectrum, an intrinsic linewidth of $\delta f = 3\,\text{kHz}$ is estimated, dictated by the phase noise of the ECL. Note that the operating point of the ECL was not optimized for minimum phase noise and that the linewidth was therefore slightly higher than the one shown in Fig. 2(d). Note also that the Kerr comb generator used here features a threshold pump power of the order of $(10...15)\,\text{mW}$ [31], [35], [36], which may be reduced further to below 5 mW [16], e.g., by increasing the quality factor of the ring resonator [16], [31], [37]. Leveraging the full 15 mW of on-chip ECL power by co-integrating the $Si_3N_4$ microresonators with the external feedback circuit of the ECL would allow to omit the EDFA and to build a fully integrated Kerr comb module similar to the one demonstrated in [16].

Our experiments show that the frequency agility of a thermally tunable ECL is already sufficient to generate a Kerr soliton frequency comb, which requires a precise control of the emission frequency to follow a tuning profile with respect to a resonance of the high-$Q$ microresonator. Note also that using independently tunable ECL as a pump offers distinct advantages over integrated Kerr comb generators that are based on self-injection locking (SIL) of the pump laser to the Kerr-nonlinear resonator [16], [37]–[42]. Specifically, pumping a Kerr comb generator by a freely tunable continuous-wave (CW) laser allows to independently control the detuning of the pump tone from the resonance of the cold ring, which is impossible for SIL. This additional degree of freedom allows to optimize the generated frequency comb, e.g., with respect to bandwidth or achievable comb-line power [39]. Specifically, single-soliton states with a large detuning and broad bandwidth may be hard to obtain in the SIL regime [42]. Note that when using a separate CW laser as a pump for Kerr-comb generation, the frequency noise of the comb lines is mainly dictated by the pump laser. A narrow-linewidth pump laser is hence crucial to achieve low-noise frequency combs.

Our demonstration further shows the importance of hybrid integration concepts, that can efficiently combine different PIC technologies in a chip-scale package. Specifically, the two $Si_3N_4$ chips used for our assemblies are based on different material platforms, namely the TriPleX® platform for the ECL filter circuit and the photonic Damascene process for the high-$Q$ microresonator. Both platforms have their own highly specific fabrication workflows that have been individually optimized and that cannot be straightforwardly combined in monolithically integrated devices on a single substrate. Using PWB or FaML for assembly of such hybrid multi-chip systems can overcome such limitations and pave the path towards efficient co-integration of TriPleX®-based ECL filter circuits with dedicated $Si_3N_4$ chips for Kerr-comb generation in a compact module, potentially using additional booster semiconductor optical amplifiers (BSOA) between the chips.

## 4. Summary

We demonstrate a hybrid integrated ECL that exploits an intra-cavity photonic wire bond (PWB) for connecting an InP reflective semiconductor optical amplifier (RSOA) to a $Si_3N_4$-based tunable external-cavity circuit. Photonic wire bonding overcomes the need for high-precision alignment during assembly, thereby permitting fully automated mass production of hybrid photonic multi-chip assemblies. The ECL offers a 90 nm tuning range (1480 nm–1570 nm) with on-chip output powers above 12 dBm and side-mode suppression ratios of up to 59 dB in the center of the tuning range. Upon optimization of the operating point, we measure an intrinsic linewidth of 979 Hz, which is among the lowest values reported for two-ring feedback architectures. The versatility of the ECL is demonstrated by using the device as narrow-linewidth tunable pump laser for Kerr frequency comb generation in a high-$Q$ $Si_3N_4$ microresonator. To the best of our knowledge, our experiments represent the first demonstration of a single-soliton Kerr comb generated with a pump that is derived from a hybrid integrated ECL. Frequency-tunable hybrid ECL can generally serve a wide range of applications from coherent communications [2] to frequency-modulated continuous-wave (FMCW) light detection and ranging (LIDAR) [12] and to optical coherence tomography [43], but the success of such concepts will finally rely on the scalability of the underlying fabrication techniques. Our experiments address this challenge and pave a path towards highly versatile hybrid light sources that combine the scalability advantages of monolithic integration with the performance and flexibility of hybrid multi-chip assemblies.

# Appendix

*A. Characterization of ECL filter circuit*

For characterization of the ECL filter circuit (TriPleX® chip), the individual racetrack resonators are first characterized by power transmission measurements in a loop-back configuration through the two auxiliary waveguide Ports 3 and 4, see Fig. 1(b). Since the measured transmission spectrum contains resonances of both racetrack resonators, we first cut out relevant data segments around each resonance and assign them to the corresponding resonator. We then analyze the dispersion relations of both racetrack resonators by extracting the frequency-dependent round-trip phase delay $\theta(\omega) = -\beta(\omega)L$ of each device, where $\beta(\omega)$ represents the frequency-dependent propagation constant of the underlying waveguides and where $L$ is the perimeter of the respective resonator, which amounts to 885.1 µm for R1 and to 857.4 µm for R2. The frequency-dependent round-trip phases of both racetrack resonators are fit by a second-order polynomial about the center frequency of our tuning range, $\omega_c = 2\pi f_c = 2\pi \cdot 196.6\,\text{THz}$ $(\lambda_c = 1525\,\text{nm})$

$$\theta(\omega) = \theta_c^{(0)} + \theta_c^{(1)}(\omega - \omega_c) + \frac{1}{2}\theta_c^{(2)}(\omega - \omega_c)^2 \qquad (1)$$

where $\theta_c^{(0)}$ corresponds to the round-trip phase delay at the center frequency $\omega_c$ whereas $\theta_c^{(1)}$ and $\theta_c^{(2)}$ give the corresponding first- and second-order derivatives. Note that $\theta_c^{(0)} = -\beta_c^{(0)}L$ is usually only know up to an integer multiple of $2\pi$ since the absolute value of the round-trip phase at the various resonances is hard to extract from the measured data. $\theta_c^{(1)} = -\beta_c^{(1)}L$ corresponds to the round-trip group delay at the center frequency $\omega_c$ and hence allows to directly extract the free spectral range (FSR) at the center frequency of the tuning range, $\Delta f_{\text{FSR}} = \Delta\omega_{\text{FSR}}/(2\pi) = 1/(2\pi \cdot \theta_c^{(1)})$, which is found to be $\Delta f_{\text{FSR,R1}} = 195.7\,\text{GHz}$ for R1 and $\Delta f_{\text{FSR,R2}} = 202.0\,\text{GHz}$ for R2. Using the relation $n_{\text{eg}} = \beta_c^{(1)}c = -\theta_c^{(1)}c/L$, we find identical refractive indices of $n_{\text{eg}} = 1.73$ for both resonators at a wavelength of $\lambda_c = 1525\,\text{nm}$ which is in good agreement with the value of $n_{\text{eg,TriPleX}} \approx 1.72$ specified for the TriPleX® platform at a wavelength of 1550 nm [26], [27]. $\theta_c^{(2)} = -\beta_c^{(2)}L$ is related to the group-delay dispersion (GDD) of the resonators and amounts to $824\,\text{fs}^2$ for R1 and to $806\,\text{fs}^2$ for R2. This can finally be related to the group-velocity dispersion (GVD) $\beta_c^{(2)}$ of the resonators, leading to $931\,\text{fs}^2/\text{mm}$ for R1 and to $940\,\text{fs}^2/\text{mm}$ for R2.

In a next step, we use the fitted dispersion relation $\theta(\omega)$ according to Eq. (1) in the through-port model of a single ring resonator with a pair of identical lossless coupling sections [44],

$$T_{\text{thrgh}}(\omega) = \left|\frac{\tau - a\tau e^{j\theta(\omega)}}{1 - a\tau^2 e^{j\theta(\omega)}}\right|^2, \qquad (2)$$

and fit it to the frequency-dependent power transmission through R1 and R2 measured in the vicinity of each resonance. In Eq. (2), $\tau$ is the real-valued amplitude transmission coefficient describing the transmission through each of the two coupling sections of the racetrack resonator, and $a$ is the real-valued amplitude transmission factor representing the round-trip loss, see [1] for a more detailed explanation. From fitting the model in Eq. (2) to our measurement data, we obtain parameters of $a \approx 0.999$ and $\tau \approx 0.928$, leading to an amplitude coupling coefficient of $\kappa = \sqrt{1-\tau^2} \approx 0.372$. The associated measurement data and fit curve are shown for one exemplary resonance dip of R1 in Fig. 4(a). Using the extracted values for the parameters $n_{\text{eg}}$, $\tau$ and $a$, we can estimate loaded $Q$-factors of approximately 20 000 at a wavelength of 1550 nm for each of the racetrack resonators R1 and R2 using the relation [45]

$$Q \approx \frac{n_{\text{eg}}\omega L}{2c}\frac{\sqrt{a\tau}}{1-a\tau^2}. \qquad (3)$$

With the extracted parameters of both resonators at hand, we can now simulate the response of the filter circuit, which comprises the Sagnac loop mirror with the Vernier pair of tunable racetrack resonators. To this end, we first calculate the frequency-dependent drop-port power transmission $T_{\text{drp}}(\omega)$ of each individual resonator, which is given by

$$T_{\text{drp}}(\omega) = \left|\frac{-\sqrt{a}\kappa^2 e^{j\theta(\omega)/2}}{1 - a\tau^2 e^{j\theta(\omega)}}\right|^2. \qquad (4)$$

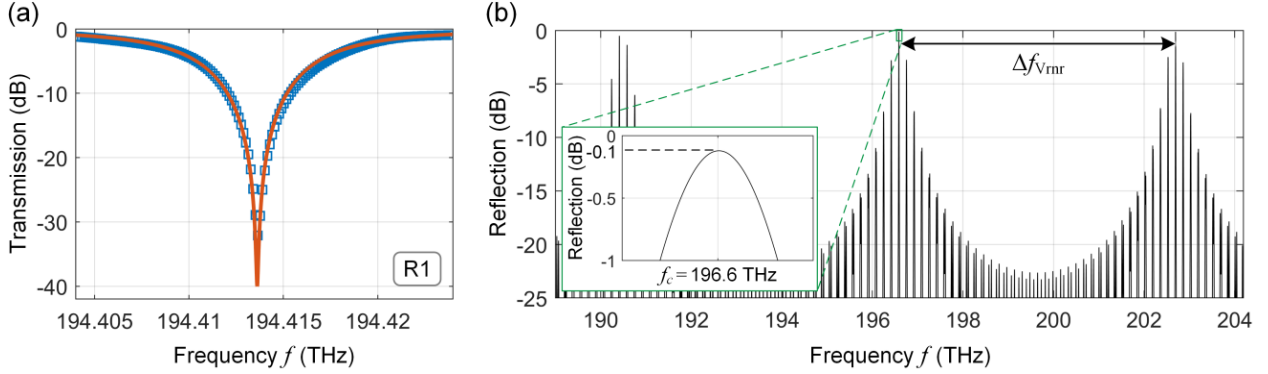

**Fig. 4.** Characterization results of the ECL filter chip containing a Sagnac loop mirror with a Vernier pair of tunable racetrack resonators R1 and R2, see Fig. 1(b). **(a)** Measured through-port power transmission $T_{\text{thrgh}}(\omega)$ (blue squares) and fit according to Eq. (2) (orange line), shown for one exemplary resonance dip of R1. **(b)** Overall frequency-dependent reflection of the Sagnac loop mirror resulting from the multiplication of the individual drop-port power transmissions of R1 and R2. The peak at the common resonance $f_c = 196.6\,\text{THz}$ indicates the on-chip reflection loss of the tunable mirror and amounts to 0.1 dB, see zoom-in.

The overall frequency-dependent reflection of the Sagnac loop mirror finally results from the multiplication of the individual drop-port power transmissions [1] and is shown in Fig. 4(b). For the plot, we assume that the racetrack resonators R1 and R2 are tuned to maximum transmission at the center frequency of our tuning range, $f_c = \omega_c/(2\pi) = 196.6\,\text{THz}$ corresponding to $\lambda_c = 1525\,\text{nm}$, such that the calculated resonances coincide at this wavelength. The peak at the common resonance indicates the on-chip reflection loss of the tunable mirror, which amounts to 0.1 dB, see zoom-in in Fig. 4(b), and is assumed to be essentially constant over the entire tuning range. From the depicted reflection spectrum, we can further estimate the spectral spacing $\Delta f_{\text{Vrnr}}$ between the strong main reflection peak and the two most prominent side peaks that are most prone to generate unwanted lasing modes, see Fig. 4(b). For our device we find $\Delta f_{\text{Vrnr}} \approx 6.2\,\text{THz}$, corresponding to $\Delta \lambda_{\text{Vrnr}} \approx 48\,\text{nm}$ when tuning the main peak to the center wavelength of $\lambda_c = 1525\,\text{nm}$, in accordance with the relation $\Delta f_{\text{Vrnr}} = \left(\Delta f_{\text{FSR,R1}} \Delta f_{\text{FSR,R2}}\right)/\left(\Delta f_{\text{FSR,R2}} - \Delta f_{\text{FSR,R1}}\right) \approx 6.3\,\text{THz}$ [1]. Note that, in contrast to earlier models of Vernier feedback structures [1], [4], [5], [8], [11], [14], the consideration of GVD in Eq. (4) leads to an asymmetric reflection spectrum. When tuned to the center wavelength of the emission range, $\lambda_c = 1525\,\text{nm}$, we find the low-frequency side peak of the reflection spectrum around $f_c - \Delta f_{\text{Vrnr}}$ to be 0.4 dB lower than the main peak, whereas the high-frequency side peak around $f_c + \Delta f_{\text{Vrnr}}$ is only 0.01 dB lower than the main peak. Notably, while the spectral spacing $\Delta f_{\text{Vrnr}}$ of the Vernier peaks is generally expected to limit the tuning range of the ECL, we find that our device features a tuning range for single-mode lasing of 11.6 THz (90 nm), which significantly exceeds the measured spacing of the Vernier peaks, $\Delta f_{\text{Vrnr}} \approx 6.2\,\text{THz}$ $\left(\Delta \lambda_{\text{Vrnr}} \approx 48\,\text{nm}\right)$. This is consistent with previous demonstrations using the same TriPleX® filter chips [9], [10]. We attribute this effect to the fact that the cavity phase tuner (CPT) gives an additional degree of freedom to favor a certain desired longitudinal lasing mode of the overall cavity, while suppressing lasing at unwanted modes in the Vernier side peaks.

*B. Photonic wire bond loss estimation*

For estimating the PWB loss, we compare the on-chip output power $P_{\text{out}}$ of the ECL to the saturated output power $P_{\text{sat}}$ of the RSOA at the same wavelength $\lambda$ and at the same injection currents $I$. As a first step, we use the data obtained from the RSOA characterization prior to assembly of the ECL module and extract the current-dependent saturated RSOA output power $P_{\text{sat}}(I)$. To this end, we launch light at a wavelength $\lambda \approx 1550\,\text{nm}$ into the RSOA through a circulator using an AR-coated lensed single-mode fiber (SMF) and measure the amplified signal after the circulator. The RSOA is similar to the one used in [1], where the details of the measurement technique are described in the Supplementary Information. In our measurements, we vary the RSOA pump current between 20 mA and 100 mA in steps of 20 mA and derive the associated relationships between the RSOA input and output power, both measured at the RSOA chip facet, see Fig. 5(a). The device was tested with 12 discrete input power levels – for better visibility, we indicated the full set of measurement points as individual dots only at a pump current of $I = 20\,\text{mA}$. The saturated output power for each pump current is found at an on-chip input power of 2.2 dBm, indicated by the respective measurement points. The horizontal dashed line indicates the saturation power $P_{\text{sat}}(100\,\text{mA}) = 11.4\,\text{dBm}$ at the RSOA facet.

To find the on-chip output power $P_{\text{out}}$ of the ECL, we couple the light emitted from the $\text{Si}_3\text{N}_4$ output WG into an SMF. We then use the fiber-chip coupling loss, obtained from a straight reference WG on the same $\text{Si}_3\text{N}_4$ chip, to estimate the on-chip power levels $P_{\text{out}}$. In this experiment, we again tune the emission wavelength to approximately $\lambda = 1550\,\text{nm}$ and again vary the pump current between 20 mA and 100 mA in steps of 20 mA. For each current, we record the output spectrum using an OSA, and extract the peak powers $P_{\text{out}}(I)$. For an injection current of $I = 100\,\text{mA}$ an on-chip output power $P_{\text{out}}(100\,\text{mA}) = 9.6\,\text{dBm}$ is measured, see Fig. 5(b) for an exemplary ECL spectrum. Taking into account the attenuation of 0.1 dB of the resonance-aligned racetrack resonators and neglecting further on-chip losses, we find the output power $P_{\text{out,PWB}}(I)$ of the PWB. The PWB coupling efficiency is then estimated by taking the previously measured saturation RSOA output power $P_{\text{sat}}(I)$ as an input from the RSOA into the PWB in Fig. 1(b), $\eta_{\text{PWB}} = P_{\text{out,PWB}}(I)/P_{\text{sat}}(I)$. This leads to an estimated insertion loss of the PWB of $a_{\text{PWB}} = -10\log_{10}(\eta_{\text{PWB}}) = (1.6 \pm 0.2)\,\text{dB}$, see Fig. 5(c).

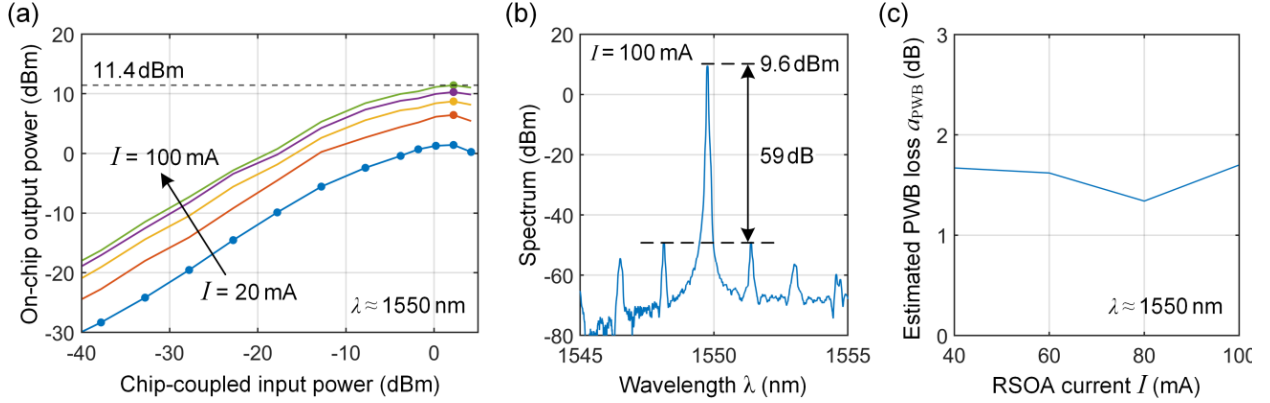

**Fig. 5.** Procedure for PWB loss estimation. We compare the on-chip output power $P_{\text{out}}$ of the ECL to the saturated output power $P_{\text{sat}}$ of the RSOA at the same wavelength $\lambda \approx 1550\,\text{nm}$ and the same injection currents $I$. (**a**) RSOA characterization results describing the current-dependent saturated output power $P_{\text{sat}}(I)$ of the RSOA. The data was obtained by launching light at a wavelength $\lambda \approx 1550\,\text{nm}$ into the RSOA through a circulator using an AR-coated lensed single-mode fiber (SMF) and by measuring the amplified signal after the circulator. We vary the RSOA pump current between 20 mA and 100 mA in steps of 20 mA and derive the associated relationships between the RSOA input and output power, both measured at the RSOA chip facet. The device was tested with 12 discrete input power levels – for better visibility, we indicated the full set of measurement points as individual dots only at a pump current of $I = 20\,\text{mA}$. The saturated output power for each pump current is found at an on-chip input power of 2.2 dBm, indicated by the respective measurement points. The horizontal dashed line indicates the saturation power $P_{\text{sat}}(100\,\text{mA}) = 11.4\,\text{dBm}$ at the RSOA facet. (**b**) Spectrum of the ECL emission when tuning the device to a wavelength of approximately $\lambda = 1550\,\text{nm}$, indicating an SMSR of 59 dB. The injection current of $I = 100\,\text{mA}$ leads to an on-chip output power of $P_{\text{out}}(100\,\text{mA}) = 9.6\,\text{dBm}$. We again vary the pump current between 20 mA and 100 mA in steps of 20 mA and extract the peak powers $P_{\text{out}}(I)$ from the spectra measured by an OSA. (**c**) Result of PWB loss estimation for different RSOA injection currents. Taking into account the attenuation of 0.1 dB of the resonance-aligned racetrack resonators and neglecting further on-chip losses, we find the output power $P_{\text{out,PWB}}(I)$ of the PWB. The PWB coupling efficiency is then estimated by taking the previously measured saturation RSOA output power $P_{\text{sat}}(I)$ as an input from the RSOA into the PWB in Fig. 1(b), $\eta_{\text{PWB}} = P_{\text{out,PWB}}(I)/P_{\text{sat}}(I)$. This leads to an estimated insertion loss of the PWB of $a_{\text{PWB}} = -10\log_{10}(\eta_{\text{PWB}}) = (1.6 \pm 0.2)\,\text{dB}$.

### C. Characterization of the Si$_3$N$_4$ chip in the Kerr comb module (CM)

The Si$_3$N$_4$ chip with the high-$Q$ ring resonator used for Kerr comb generation in the second part of the experiment was fabricated using the photonic Damascene reflow process [31], [32]. The waveguide cross section has a width of 2.2 µm and a height of 900 nm to ensure anomalous group-velocity dispersion (GVD) as needed for generation of soliton Kerr combs. The EC located at the input and output facets rely on waveguide structures in a second Si$_3$N$_4$ layer, which is only 200 nm thick, and which is spaced from the 900 nm-thick layer by a 100 nm-thick SiO$_2$ layer – similar structures were also used in [46]. The waveguides in the 900 nm-thick layer are down-tapered over a length of 300 µm from the usual width of 2.2 µm to a final width of 200 nm at a distance of 300 µm from the chip edge. Over the length of this tapered section, the waveguides in the 200 nm-thick upper Si$_3$N$_4$ layer are simultaneously up-tapered from an initial width of 200 nm to 2.1 µm and are then down-tapered again to a final width of 300 nm at the chip facet. This leads to a slightly elliptical mode field with a measured 1/e²-width of 3.8 µm in the horizontal and of 3.5 µm in the vertical direction, which is coupled to a tapered PWB with initial cross-section of 4.6 µm × 4.0 µm.

For characterization of the Si$_3$N$_4$ chip, we employ frequency comb-calibrated laser spectroscopy [47], [48] for an exact measurement of the dispersion characteristics and the quality factor of R3. All characterization measurements and all comb-generation experiments are performed using the fundamental TE mode of the on-chip waveguides. The dispersion can again be extracted from the transmission spectrum. In contrast to the model in Eq. (1), the dispersion of the ring resonators for Kerr-comb generation is usually represented in terms of the deviation $D_{\text{int}}$ of the resonance frequencies from an equidistant grid, where the various resonant frequencies of R3 are given by $\omega_n$, and where the index $n = 0$ corresponds to a central resonance frequency $\omega_0 = 2\pi \cdot 192.3\,\text{THz}$,

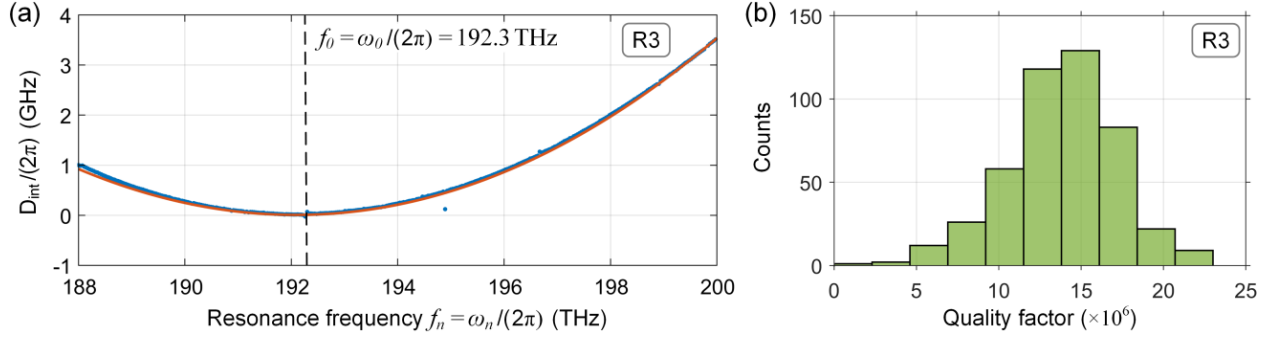

**Fig. 6.** Characterization results of the high-$Q$ $Si_3N_4$ ring resonator R3 used for Kerr-comb generation. **(a)** Measured dispersion landscape [47] $D_{int}/(2\pi)$ (blue dots) and fit according to Eq. (5) (orange line), revealing an FSR of $D_1/(2\pi) = 35.4\,\text{GHz}$ and a second-order dispersion element $D_2/(2\pi) = 139.6\,\text{kHz} > 0$, corresponding to anomalous group velocity dispersion as needed for Kerr-comb generation. **(b)** Results of a statistical analysis of the various resonances of R3, revealing an average quality factor of $Q \approx 15 \times 10^6$.

$$D_{int} = \omega_n - (\omega_0 + nD_1) = \sum_{m \geq 2} \frac{n^m D_m}{m!}. \quad (5)$$

In this representation, the spacing of the resonance frequencies $\omega_n$ is given by $D_1$, corresponding to the reciprocal of $\theta_c^{(1)}$, i.e., $D_1 = \Delta\omega_{FSR} = 1/\theta_c^{(1)}$, whereas $D_2$ represents the GVD. A measured dispersion landscape for ring resonator R3 is given in Fig. 6(a). Fitting Eq. (5) to the measured data, we can find an FSR of $D_1/(2\pi) = 35.4\,\text{GHz}$ as well as a positive second-order dispersion element $D_2/(2\pi) = 139.6\,\text{kHz} > 0$, indicating anomalous GVD as needed for Kerr-comb generation. To estimate the quality factor of the ring resonator R3, each resonance of the measured transmission spectrum is fitted using a model based on coupled-mode theory [49], [50]. The results of a statistical analysis of the resonator transmission spectra is shown in Fig. 6(b), revealing an average quality factor of $Q \approx 15 \times 10^6$.

**Funding.** This work was supported by the Deutsche Forschungsgemeinschaft (DFG, German Research Foundation) under Germany's Excellence Strategy via the Excellence Cluster 3D Matter Made to Order (EXC-2082/1 – 390761711) and by the DFG Collaborative Research Center WavePhenomena (SFB 1173, Project-ID 258734477), by the EU Horizon 2020 Marie Skłodowska-Curie Innovative Training Network MICROCOMB (#812818), by the EU Horizon 2020 research and innovation programme TeraSlice (#863322), by the BMBF projects Open6GHub (#16KISK010), OPALID (#13N14589), STARFALL (#16KIS1420) and DiFeMiS (#16ES0948), the latter being part of the programme "Forschungslabore Mikroelektronik Deutschland" (ForLab), by the ERC Consolidator Grant TeraSHAPE (#773248), by the Alfried Krupp von Bohlen und Halbach Foundation, and by the Karlsruhe School of Optics & Photonics (KSOP).

**Acknowledgments.** We thank Ute Troppenz and Martin Möhrle (Fraunhofer Heinrich-Hertz-Institut (HHI), Germany) for providing the InP-based components.

**Disclosures.** P.-I.D. and C.K. are co-founders and shareholders of Vanguard Photonics GmbH and Vanguard Automation GmbH, companies engaged in exploiting 3D nano-printing in the field of photonic integration and assembly. C.K. and T.J.K. are co-founders and shareholders of Deeplight S.A., a start-up company engaged in the development of advanced light sources such as tunable lasers and frequency comb generators. P.M., Y.X., M.B., P.-I.D., and C.K. are co-inventors of patents owned by Karlsruhe Institute of Technology (KIT) in the technical field of the publication. P.-I.D. is now an employee of Vanguard Automation GmbH. M.B. and Y.X. are now employees of Nanoscribe GmbH, a company selling 3D lithography systems. D.G. is now an employee of Chilas B.V., a company selling narrow-linewidth tunable external-cavity lasers. The other authors Y.C., Y.B., R.D., J.L., H.P., S.R. and W.F. declare no conflict of interest.

**Data availability.** Data underlying the results presented in this paper are not publicly available at this time but may be obtained from the authors upon reasonable request.